\documentstyle[12pt]{article}

\def\hybrid{\topmargin -20pt  \oddsidemargin 0pt
      \headheight 0pt   \headsep 0pt
      \textwidth 6.25in 
      \textheight 9.5in 
      \marginparwidth .875in
      \parskip 5pt plus 1pt   \jot = 1.5ex}

\hybrid

\def\x{\times}

\def\o+{\oplus}
\def\ra{\rightarrow}

\def\beqa{\begin{eqnarray}}
\def\eeqa{\end{eqnarray}}
                       
\newcommand{\ov}{\overline}

\newcommand{\la}{\lambda}
\newcommand{\La}{\Lambda}
\newcommand{\si}{\sigma}

\newcommand{\resetcounter}{\setcounter{equation}{0}}

\begin{document}
\thispagestyle{empty}
\rightline{LMU-ASC 40/11}
\vspace{2truecm}
\centerline{\bf \LARGE Higgs Multiplets}
\vspace{.3truecm}
\centerline{\bf \LARGE in Heterotic GUT Models}

\vspace{1.5truecm}
\centerline{Gottfried Curio\footnote{gottfried.curio@physik.uni-muenchen.de; 
supported by DFG grant CU 191/1-1}} 

\vspace{.6truecm}

\centerline{{\em Arnold-Sommerfeld-Center 
for Theoretical Physics}}
\centerline{{\em Department f\"ur Physik, 
Ludwig-Maximilians-Universit\"at M\"unchen}}
\centerline{{\em Theresienstr. 37, 80333 M\"unchen, Germany}}

\vspace{1.0truecm}

\begin{abstract}
For supersymmetric GUT models from heterotic string theory, built from 
a stable holomorphic $SU(n)$ vector bundle $V$ on a Calabi-Yau threefold $X$,
the net amount of chiral matter can be computed by a Chern class
computation. Corresponding computations for the number $N_H$ of Higgses
lead for the phenomenologically relevant cases of GUT group $SU(5)$ or $SO(10)$
to consideration of the bundle $\La^2 V$. In a class of bundles where everything can be 
computed explicitly (spectral bundles on elliptic $X$) we find that the computation 
for $N_H$ gives a result which is in conflict with expectations.
We argue that this discrepancy has its origin in the subtle geometry of the spectral data 
for $\La^2 V$ and that taking this subtlety into account properly 
should resolve the problem.
\end{abstract}

\newpage

\section{Introduction}

For supersymmetric GUT models from heterotic string theory, built from 
a stable holomorphic vector bundle $V$ on a Calabi-Yau threefold $X$,
the net amount of chiral matter can be computed by a Chern class
computation. Corresponding computations for the number $N_H$ of Higgses
lead for the phenomenologically relevant cases of GUT group $SU(5)$ or $SO(10)$
to consideration of the bundle $\La^2 V$. In a class of bundles where everything can be 
computed explicitly (spectral bundles on elliptic $X$) we find that the computation
for $N_H$ gives a result which is in conflict with expectations.
We argue that this discrepancy has its origin in the subtle geometry of the spectral data 
for $\La^2 V$ and that taking this subtlety into account properly 
should resolve the problem.

We recapitulate the standard field theory situation and the
corresponding heterotic string version in {\em sect.~\ref{field and string theory}}. 
In {\em sect.~\ref{mathematics}} we 
develop the specific heterotic string theory model we employ 
and identify in {\em sect.~\ref{puzzle}}
the puzzling correction term to the expected identity; 
we also show the appearance of precisely
this correction already in an analogous auxiliary problem given 
by an Euler number computation in a
ramified covering (we later argue that because of the coincidence 
of the complicated inner structure 
of this correction term in both problems, the original one and the reduced one, 
our solution of the 
reduced problem suggests that the essential content of the original problem 
is already captured); 
further we give a first discussion of the individual cases $n=4, 5$
and argue for the necessity to consider also the general case $n\geq 6$. 
The resolution of the puzzle in the reduced auxiliary problem for this general case 
is given in {\em sect.~\ref{resolution}};
in the light of this general discussion the cases $n=4, 5$ are investigated again in 
greater detail. We conclude in {\em sect.~\ref{Conclusion}}.

\section{\label{field and string theory}
Matter multiplets in susy GUT's in field and string theory}

For convenience of the reader and to establish some notation we start in assembling 
some standard facts about matter multiplets in susy GUT's in field and string theory.

\subsection{\label{field theory}The field theory: matter and Higgs multiplets in susy GUT's}

We consider supersymmetric grand unified extensions of the standard model ($SM$),
more precisely theories with one of the GUT groups $SU(5)$ or  $SO(10)$ (or $E_6$).

\subsubsection{The standard model}

Let us begin by recalling some well-known facts.
For the $SM$ itself one has the fermionic matter 
content
\beqa
SM \; fermions &=& Q \oplus \ov{u}\oplus \ov{d}\oplus L\oplus \ov{e}\\
&=&({\bf 3}, {\bf 2})_{1/3}\oplus 
({\bf \ov{3}}, {\bf 1})_{-4/3}\oplus ({\bf \ov{3}}, {\bf 1})_{2/3}\oplus 
({\bf 1}, {\bf 2})_{-1}\oplus
({\bf 1}, {\bf 1})_2
\eeqa
(indicating the decomposition 
under the $SM$ gauge group $SU(3)_c\x SU(2)_{ew}\x U(1)_Y$); one has
the anti-particles as well and a net amount of $3$ chiral families
(besides these unpaired multiplets which appear at low energies 
the remaining conjugate pairs pair up with high scale masses).
Furthermore, in the supersymmetric extension of the standard model (which we consider)
the electro-weak Higgs doublet comes as a conjugate pair: 
$H_u=({\bf 1}, {\bf 2})_{1}, H_d=({\bf 1}, {\bf \ov{2}})_{-1}$;
the mass term $\mu$ sits at the weak scale.

\subsubsection{The $SU(5)$ GUT theory}

For the GUT group $SU(5)$ the fermionic matter sits in
\beqa
{\bf \ov{5}}&=&L\oplus \ov{d}\\
{\bf 10}&=&Q\oplus \ov{u}\oplus \ov{e}
\eeqa
(indicating the decomposition under the SM subgroup). The Higgs sits in
a ${\bf 5}$ and under a breaking of $SU(5)$ to the $SM$ gauge group one has
to understand the splitting between the electro-weak doublet which has to 
sit at the weak scale and a remaining color triplet which is superheavy.
One gets the conditions 
(with the net numbers $N_{\bf R}:=n_{{\bf R}}-n_{{\bf \ov{R}}}$)
\beqa
N_{{\bf \ov{5}}}&=& N_{{\bf 10}}\; = \; 3\\
\label{Higgs SU5}
n_{\bf 5}&\geq &1
\eeqa
If one has a Wilson line breaking to the $SM$ one may pose instead of (\ref{Higgs SU5}) 
the stronger condition of having just {\em one} $SM$ Higgs doublet conjugate (vector-like) pair.

\subsubsection{The $SO(10)$ GUT theory}

For $SO(10)$ the previous fermionic matter content combines together with a right-handed
neutrino (which is a singlet under the $SM$ gauge group) to the chiral spinor
representation 
\beqa
{\bf 16}&=&{\bf \ov{5}}\oplus {\bf 10} \oplus {\bf 1}
\eeqa
(indicating the decomposition under $SU(5)$) whereas a conjugate pair of ${\bf 5}$'s of $SU(5)$ 
(containing a Higgs conjugate pair) combines to a ${\bf 10}$ of $SO(10)$ 
(note that this is a real representation). One gets the conditions 
\beqa
N_{{\bf 16}}&=&3\\
\label{Higgs SO10}
n_{{\bf 10}}&\geq &1
\eeqa
If one has a Wilson line breaking to the $SM$ 
one may pose instead of (\ref{Higgs SO10}) 
the stronger condition of having just {\em one} 
$SM$ Higgs doublet conjugate (vector-like) pair.
(Ideally taking the invariant subspace projects out the unwanted colour triplets.)

(In the next step ${\bf 16}$ and ${\bf 10}$ would combine with a singlet of $SO(10)$ to
the fundamental representation ${\bf 27}$ of $E_6$. This case turns out 
to be not interesting for us, cf.~below.)

\subsection{\label{string theory}The string theory: heterotic models}

A supersymmetric heterotic string model in four dimensions arises as
the low energy effective theory from a compactification of the tendimensional heterotic string 
on a Calabi-Yau threefold $X$ endowed with a polystable holomorphic vector bundle $V'$. 
Usually one takes $V'=(V, V_{hid})$ with
$V$ a {\em stable} bundle considered to be embedded in (the visible) $E_8$ 
($V_{hid}$ plays the corresponding role for the second hidden $E_8$); the
commutant of $V$ in $E_8$ gives the unbroken gauge group in four dimensions.
(Though it is not essential we restrict our attention to $V$.)
The $\mu$ term will arise as a moduli vev in a cubic interaction $\phi H_u H_d$.

One has in the decomposition of the tendimensional gauge group
with respect to $G\x H_V$ (where $G$ is one of the 
fourdimensional GUT gauge groups $E_6, SO(10)$ and $SU(5)$ 
and $H_V$ the corresponding structure group $SU(n)$, for $n=3,4,5$,
of the bundle $V$) besides the fourdimensional gauge group 
$({\bf ad}, {\bf 1})$ and the singlets $({\bf 1}, {\bf ad})$
(these two terms are indicated below by "...") the following matter multiplets
(with ${\bf 6}=\La^2 {\bf 4}$ and ${\bf \ov{10}}=\La^2 {\bf \ov{5}}$)
\beqa
{\bf 248}&=&({\bf 16}, {\bf 4})\oplus c.c.
\oplus ({\bf 10}, {\bf 6}) \oplus \dots
\;\;\;\;\;\;\;\;\;\;\;\;\;\;\;\;\;\; \mbox{for}\; n=4\\
&=& ({\bf \ov{5}}, {\bf \ov{10}})\oplus c.c. \oplus 
({\bf 10}, {\bf \ov{5}})\oplus c.c. \oplus \dots 
\;\;\;\;\;\;\;\;\; \mbox{for}\; n=5
\eeqa
(for $n=3$ one has just $({\bf 27}, {\bf 3})\oplus c.c.\oplus \dots$;
$\La^2 \, {\bf 3}$ is then no new representation of $H_V$ but rather ${\bf \ov{3}}$).
This leads to the number of matter multiplets as given 
by\footnote{\label{X cohomology}all cohomology groups are taken over $X$ 
unless indicated otherwise}
\beqa
n_{\bf 16}=h^1(V),\;\;\; n_{\bf 10}=h^1(\Lambda^2 V) 
\;\;\;\;\;\;\;\;\;\;\;\;\;\;\;\;\;\;\;\;\;\;\;\;\;\;\;\;\;\; 
\mbox{for}\; n=4\\
n_{{\bf \ov{5}}}=h^1(\Lambda^2 \ov{V}),\;\;\; n_{{\bf 10}}=h^1(\ov{V})
\;\;\;\;\;\;\;\;\;\;\;\;\;\;\;\;\;\;\;\;\;\;\;\;\;\;\;\;\;\;\;\;
\mbox{for} \; n=5
\eeqa
Note that in both cases the number of Higgses is related to a computation
of $h^1(\La^2 V)$. This is the rationale for the title of this note and 
decisive for what we will do in it.

For the net amount of chiral matter one has
\beqa
N_{{\bf 16}}&=&-\frac{1}{2}c_3(V)
\;\;\;\;\;\;\;\;\;\;\;\;\;\;\;\;\;\;\;\;\;\;\;\;\;\;\;\;\;\;\;\;\;\;
\;\;\;\;\;\;\;\;\;\;\;\;\;\;\;\;\;\;\;\;\;\;\;\;\;\;\;\;\;\;\;\;\;\;\;\;\;
\mbox{for}\; n=4\;\\
N_{{\bf 10}}&=&+\frac{1}{2}c_3(V) \; , \; 
N_{{\bf \ov{5}}}\; =\; +\frac{1}{2}c_3(\La^2 V)\; =\; +\frac{1}{2}(n-4)c_3(V)
\;\;\;\;\;\;
\mbox{for}\; n=5\;
\eeqa
Here the case of the self-conjugate (real) ${\bf 10}$ of $SO(10)$ does not occur, of course.

For example, for $n=5$ with Wilson line breaking to the $SM$
one may get as low-energy particle spectrum
(in a case of absence of exotics, i.e.~where $n^{\pm}_{\bf \ov{10}}=0$ and $n_{{\bf 5}}^+ =0$)
\beqa
\label{first line}
n_{\bf 10}^+ =3 &,& n_{\bf 10}^- =3, \\
\label{second line}
n_{{\bf \ov{5}}}^+ =3 &,& n_{{\bf \ov{5}}}^- =3+1\\
\label{third line}
n_{{\bf 5}}^+ =0 &,& n_{{\bf 5}}^- =1
\eeqa
(where $\pm$ indicates the weight under the ${\bf Z_2}$ Wilson line action). 
Here (\ref{first line}) gives $\ov{u}\oplus \ov{e}$ and $Q$,
(\ref{second line}) $\ov{d}$ and $L$ together ("+") with $H_d$ and
(\ref{third line}) finally $H_u$.

\section{\label{mathematics}Specific heterotic models: 
spectral cover bundles on elliptic Calabi-Yau threefolds}

\resetcounter

Now we consider a specific class of heterotic models: those where $X$
carries an elliptic fibration $\pi:X\ra B$ with section $\si$ and 
$V$ is given by the spectral cover construction [\ref{FMW}].

\subsection{The bundle $V$ itself}

So we start with a $n$-fold ramified cover surface $C_V$ of 
class\footnote{\label{class notation}and of equation $w=a_0+a_2x+a_3y+a_4x^2+a_5xy$
in Weierstrass coordinates for $n=5$ (and with $a_5=0$ for $n=4$), cf.~[\ref{FMW}];
by abuse of notation we will, 
here and in all similar cases yet to come,
denote the cohomology class $[M]$ of a geometric variety $M$ also just by $M$} 
\beqa
[C_V]&=&n\si + \eta
\eeqa 
(over the embedded base surface\footnote{\label{standard bases}The standard examples are the 
Hirzebruch surfaces ${\bf F_k}$ ($k=0,1,2$), the del Pezzo surfaces
${\bf dP_k}$ ($k=0, \dots, 8$) and the Enriques surface.} 
$B$ of class $\si$) 
where a line bundle $L_V$ is given with
\beqa
c_1(L_V)&=&
\Bigg[ \frac{n\si+\eta+c_1}{2}+\la\Big(n\si-(\eta-nc_1)\Big)\Bigg]\Bigg|_{C_V}
\eeqa
(where $c_1:=c_1(B)$, $\eta\in H^2(B, {\bf Z})$ and all pull-backs
by $\pi$ here and below are suppressed).
An important role is played by the intersection curve $A_V:=C_V\cap \si$
which can be considered to lie either in $C_V$ or in $\si=\si(B)=B$:
one knows that the first cohomology of the bundle $V$ is localised along the curve in $B$ 
where one of the $n$ line bundle summands 
(into which $V$ decomposes when restricted to a fibre) 
becomes trivial, i.e.~where one of the $n$ fibre points of the cover surface $C_V$ meets $\si$; 
this leads to the curve (with corresponding cohomology class$^{\ref{class notation}}$)
\beqa
A_V &=&C_V\cap B\\
{[}A_V{]} &=&\eta-nc_1
\eeqa

\subsubsection{Consideration of the ramification locus}

Let us denote by $r_V$ the ramification locus (above in $C_V$) of $\pi: C_V\ra B$
(as usual $r_V$ will denote also the cohomology class).
From $K_{C_V}=\pi_{C_V}^* K_B + r_V$ one finds 
\beqa
r_V&=&n\si+ \eta + c_1 \Big|_{C_V}\\
&=&n\Big[ 2\eta-(n-1)c_1\Big] \si + \eta(\eta+c_1)
\eeqa
This is the class of the curve on $C_V$ of ramification points (above), i.e.~of
points where in a fibre some points coincide: $p_i=p_j$ ($i\neq j$). 
For the class of self-mirror points among these doubly counting points of $C_V$ 
(if $p_i+p_j=0$, or equivalently $p_i=\tau p_j$
under the standard involution $\tau$ [\ref{FMW}], we call the pair $(p_i, p_j)$ 
$\tau$-conjugate or a mirror pair)
one finds
\beqa
r_V\cdot \si_{II}&=&4 \eta^2 - 2(n-2)\eta c_1 + n (n-1) c_1^2
\eeqa
(where $\si_{II}=\si+\si_2$ denotes the fourfold section of two-torsion points 
and $\si_2$ is the trisection of {\em strict} two-torsion points; the two components of $\si_{II}$
being disjoint one gets for $\si_2$ the class $3(\si+c_1)$).

\subsection{The derived bundle $\La^2 V$}

The bundle $\La^2 V$ is again stable and spectral [\ref{DHOR}] (we assume $n\geq 3$ from now on), 
its cover surface $C_{\La^2 V}$ has\footnote{this follows with
$c_2(\La^2 V)=(n-2)c_2(V)$ from the general representation of $c_2(U)$ as $\eta_U \si + \omega$
for any spectral bundle $U$ with class $C_U=rk(U)\si + \eta_U$ of its cover surface $C_U$}
class 
\beqa
[C_{\La^2 V}]&=&\frac{n(n-1)}{2}\si + (n-2)\eta
\eeqa

Similarly as for the original bundle $V$ one realises, 
when considering $\La^2 V$ just as another spectral bundle in itself, 
that its first cohomology is localised along the 
curve\footnote{Note that the localization curve $A_V$ for $V$ 
(where the matter computations take place)
can be 'turned off' discretely by an appropriate choice of $\eta$:
just take $\eta-nc_1=0$ (which is of course still effective); so for this choice
$c_3(V)$ should vanish - as it does indeed. If one tries to apply the same
logic to the localization curve $A_{\La^2 V}$ of $\La^2 V$, however,
one realises that the corresponding choice $\eta=\frac{n(n-1)}{2(n-2)}c_1$
(let us assume this would be an integral class, say for $n=3$ or $4$)
would violate the demand that $\eta-nc_1$ should be effective;
thus the analogous procedure is not legitimate there, which is good
as $c_3(\La^2 V)$, being proportional to $c_3(V)$, does {\em not} have
a cohomological factor $[A_{\La^2 V}]$ but still just $[A_V]$.
(This remark is relevant for $n=4$; for $n=3$ the two choices of $\eta$,
and already $[A_V]$ and $[A_{\La^2 V}]$ are identical and so no contradiction arises there
(of course, for $n=3$ already $\La^2 V$ itself is just $\ov{V}$).)}
\beqa
\label{new curve}
A_{\La^2 V}&=&C_{\La^2 V}\cap B\\
{[}A_{\La^2 V}{]}&=&(n-2)\eta-\frac{n(n-1)}{2}c_1
\eeqa
We assume from now on that $c_1$ is 
effective (thus excluding among the standard bases just the Enriques surface, 
cf.~footn.~\ref{standard bases})
such that with $A_V$ of class $\eta-nc_1$ also $(n-2)\eta-n(n-2)c_1$ is effective 
and thus also (\ref{new curve}).

One can also, however, consider this locus from the perspective of the
original bundle $V$, from which $\La^2 V$ is derived, and {\em its}
spectral cover surface $C_V$. Then one finds
that $H^1(\La^2 V)$ is localised along the curve of points $b$ in $B$
where in the corresponding fibre $C_b=\pi^{-1}(b)$ two 
{\em generically different} points $p_i$ and $p_j$ of $C_V$ fulfil $p_i+p_j=0$
(for, if $p_i$ are the points of $C_V$ on $C_b$, 
then the $p_i+p_j$ with $i< j$ are the points of $C_{\La^2 V}$ there)
or, equivalently, are conjugated under the standard involution $\tau$. 
Thus $A_{\La^2 V}$ has as double cover the {\em non-trivial} divisor component 
$\widetilde{A}_{\La^2 V}$ 
in $C_V\cap \tau C_V$, the other ('trivial')
components being given by $C_V\cap \si_{II}$.
Now, similarly as the surface $C_V$, an $n$-fold ramified cover of $B$, does not just have $n\si$
as its cohomology class but rather $n\si+\eta$, here
the curve $\widetilde{A}_{\La^2 V}$ has not just $2 A_{\La^2 V}\si$ as cohomology class;
rather one gets, by decomposing the intersection of $C_V$ 
with its conjugate (which has the same cohomology class),
\beqa
\widetilde{A}_{\La^2 V}&=&\Big[ C_V-\si_{II}\Big]\Big|_{C_V}=
\Big[ (n-4)\si +\eta - 3 c_1 \Big] \Big|_{C_V}\\
 &=&2 A_{\La^2 V}\si +\eta(\eta-3c_1)
\eeqa
(where the second line is to be read in $X$).
From the ramified double covering 
$\pi_C: \widetilde{A}_{\La^2 V}\ra A_{\La^2 V}$
one gets (as relation of degrees)
\beqa
\label{Euler number comparison}
\frac{1}{2}c_1(T \widetilde{A}_{\La^2 V})|_{\widetilde{A}_{\La^2 V}}&=&
c_1(T A_V)|_{A_V} - \frac{1}{2}\si_{II}\cdot \widetilde{A}_{\La^2 V}
\eeqa
(after multiplying the Euler number relation by $1/2$).

For the (degree of the) Chern class of the line bundle $L_{\La^2 V}$, 
when restricted to the corresponding intersection curve, one gets according to [\ref{DHOR}]
\beqa
\label{restricted Chern class}
c_1(L_{\La^2 V})|_{A_{\La^2 V}}&=&
\Big[ c_1(L_V)-\frac{1}{2}\si_{II} \Big]\cdot \widetilde{A}_{\La^2 V}
\eeqa

\section{\label{puzzle}The puzzle}

\resetcounter

Application of the Leray spectral sequence to the elliptic fibration
gives $H^1(X, U)\cong H^0(A_U, (L_U\otimes K_B)|_{A_U})$ where $U$ is either
$V$ or $\La^2 V$ (or $\ov{V}, \La^2 \ov{V}$).
This gives the estimate$^{\ref{X cohomology}}$
\beqa
h^1(U)&\geq & \mbox{deg} ((L_U\otimes K_B)|_{A_U}) +1-g_{A_U}\nonumber\\
&=&\Big[ c_1(L_U)-\frac{A_U+c_1}{2}\Big] \cdot A_U
\eeqa
The one term which is missing 
in our computation here (which uses the Riemann-Roch formula just as an estimate
instead of an equation) is precisely the 'other' term 
describing the anti-multiplet,
which combines (subtractively) with the estimated term to give the {\em net}
amount; this other term is
\beqa
\label{anti-matter}
h^1(\ov{U})&=&h^0(A_U, (L_U\otimes K_B)^*|_{A_U}\otimes K_{A_U})
\eeqa

The final identification just mentioned holds for $U=V$ in general, therefore one 
expects\footnote{Although there is a certain difference between $U=V$ and $U=\La^2 V$ -  
the latter has a spectral line bundle $L_{\La^2 V}$ 
which is not the restriction of a line bundle on $X$ to $C_{\La^2 V}$ -  
this is not relevant for the assumptions going in (\ref{anti-matter}) 
and thus not relevant for the point in question.} 
it to hold also for $U=\La^2 V$. 
Here one assumes that the curve $A_{\La^2 V}$ is smooth, 
reduced and irreducible.
According to [\ref{DHOR}] one has indeed that 
{\em generically} (although perhaps not in specific phenomenologically relevant
cases)
the curve $A_{\La^2 V}$ is smooth and irreducible\footnote{with "irreducible"
possibly intended to mean
reduced and irreducible as assumptions for the applicability of the ordinary
Riemann-Roch formula are at stake there}. One should note that when the authors of [\ref{DHOR}]
make computations in a {\em specific} model they do not use this assumption.

Therefore one expects
\beqa
h^1(U)-h^1(\ov{U})&=&\mbox{deg} ((L_U\otimes K_B)|_{A_U}) +\frac{1}{2}e_{A_U}\\
&=&\Big[ c_1(L_U)-\frac{A_U+c_1}{2}\Big] \cdot A_U
\eeqa
Evaluating this for $U=V$ gives $-\la \eta(\eta-nc_1)$,
thus giving, according to [\ref{C}], indeed the correct result
$-\frac{1}{2}c_3(V)$.

When doing the corresponding computation for $U=\La^2 V$ one meets a surprise, however: 
one gets (with (\ref{restricted Chern class})) for $h^1(U)-h^1(\ov{U})$, 
instead of having just the expected term 
$-\frac{1}{2}c_3(\La^2 V)=-(n-4)\frac{1}{2}c_3(V)=-(n-4)\la \eta (\eta - n c_1)$,
in addition also a correction term 
\beqa
\label{corrected bound}
\mbox{deg} ((L_{\La^2 V}\otimes K_B)|_{A_{\La^2 V}}) +\frac{1}{2}e_{A_{\La^2 V}}
&= & \Delta_n -(n-4)\la \eta (\eta - n c_1)
\eeqa
for which one has the following expression
\beqa
\Delta_n&=&-\frac{n-3}{2}\Bigg[ (n-4)\, \eta^2 
- (n^2-3n-2)\, \eta c_1
+ n(n-1)\frac{n-2}{4}\, c_1^2 \Bigg]\\
&=&\left\{ \begin{array}{ll}
(\eta-3c_1)c_1 
\;\;\;\;\;\;\;\;\;\;\;\;\;\;\;\;\;\;\;\;\;\;\;\;\;\;\;\;\;\;\;\;\;\;
\mbox{for} \;\; n=4\\
-(\eta-3c_1)(\eta-5c_1)
\;\;\;\;\;\;\;\;\;\;\;\;\;\;\;\;\;\;\;\;
\mbox{for} \;\; n=5
\end{array} \right.
\eeqa
($\Delta_3$ vanishes but there we had not to consider any $\La^2 V$; 
we assume $n\geq 4$ from now on). 

\subsection{An analogous, reduced problem}

To locate the point where an error might have come in we switch to an auxiliary problem:
as a mismatch does occur in the mentioned expressions let us check
whether at least the simpler Euler number comparison 
in (\ref{Euler number comparison}) comes out correctly.
One finds with\footnote{where the degree $\mbox{deg} Br$ of the branching locus 
below in $A_{\La^2 V}\subset B$ can also be computed above in 
$\widetilde{A}_{\La^2 V}\subset C_V$, 
namely from the ramification locus as $\widetilde{A}_{\La^2 V}\cdot \si_{II}$}
\beqa
K_{\widetilde{A}_{\La^2 V}}&=&6\Big[n-2\Big]\eta^2 
- 2\Big[3n^2-3n-4\Big]\eta c_1 +n(n-1)\Big[2n-1\Big] c_1^2\\
K_{A_{\La^2 V}}&=&\Big[n-2\Big]^2\eta^2-\Big[(n-2)(n^2-n+1)\Big]\eta c_1 
+n(n-1)\Big[\frac{n^2-n+2}{4}\Big]c_1^2\\
\mbox{deg} Br&=&4\eta^2-2\Big[n+4\Big]\eta c_1 +n(n-1)c_1^2
\eeqa
that one has, surprisingly, the relation again with a correction term
\beqa
\frac{1}{2}e_{\widetilde{A}_{\La^2 V}}&=&
e_{A_{\La^2 V}}-\frac{1}{2}\, \mbox{deg} \, Br -2 \Delta_n
\eeqa
\indent
As the same correction term (up to a constant factor) 
occurs here just in an Euler number computation
we will investigate now more closely its structure, especially in connection
to the ramification structure of double covering 
$\widetilde{A}_{\La^2 V}\ra A_{\La^2 V}$. 

The philosophy we follow from now on will be that the appearance of precisely the 
same correction term already in the analogous auxiliary problem 
(given by an Euler number computation in a ramified covering) 
does not just represent an accidental coincidence 
(the rationale for this philosophy consists in the 
complicated inner structure of this same correction term for both problems, 
the original one and the reduced one). Therefore, we argue, 
our later explanation and complete solution of the reduced problem suggests 
that the source and essential content of the original problem is already captured.

\subsection{The sign of the correction term}

For the physical application the cases $n=4$ and $5$ are, of course, 
the ones of greatest interest. However, the 'puzzle' exists
already on a purely mathematical level, and so the contradiction has to be solved for 
all $n\geq 6$ as well (for $n=3$ the correction vanishes).

Let us investigate first the sign of the correction term.
One of the assumptions in the spectral cover set-up is
that $e_n:=\eta-nc_1$, being the class of $A_V$ in $B$, 
is (the class of an) effective (divisor).
Thus we get
\beqa
\Delta_n&=&\left\{ \begin{array}{ll}
(e_4+c_1)c_1 \;\;\;\;\;\;\;\;\;\;\;\;\;\;\;\;\;\;\;\; \mbox{for} \;\; n=4\\
-(e_5+2c_1)e_5\;\;\;\;\;\;\;\;\;\;\;\;\;\;\; \mbox{for} \;\; n=5
\end{array} \right.
\eeqa
Let us assume from now on that $c_1$ is ample. This excludes among the bases
${\bf F_k}$ (with $k=0,1,2$), ${\bf dP_k}$ (with $k=0, \dots, 8$) and 
Enriques (which was already excluded) just the latter and ${\bf F_2}$.
This gives $\Delta_4>0$.
For $n=5$ one may wish to assume that $e_5$ is not only effective but even 
ample (if $C_V$ is ample then $e_n=[A_V]$ will be ample, too):
then one gets $\Delta_5<0$.

As explained above the correction issue has also to be explained for all $n\geq 6$.
Now, for $n=6$ one finds that with $\Delta_6=3\Delta_5$ the same reasoning as in the case $n=5$
applies. One could now compute the higher $\Delta_n$ individually and consider them case by case;
one finds, for example, $\Delta_8=-5(e_8+\eta+c_1)(e_8+2c_1)$. In general one has
\beqa
\Delta_n&=&-\frac{n-3}{2}\Bigg[ (n-4)e_n^2+(n^2-5n+2)e_n c_1+n\frac{(n-2)(n-5)}{4}c_1^2\Bigg]
\eeqa
One learns that for $e_n$ ample the term in big brackets here is always positive for $n\geq 5$.
We will assume from now on that $e_n=[A_V]$ is ample.
We find that 
the expression $\delta_n:=-2\Delta_n$ (the rationale for this definition will become clear
below) is always positive for $n\geq 5$.

\subsection{\label{cases 4 and 5}A preliminary look at the cases $n=4$ and $n=5$}

So, one of the assumptions going in the derivation of (\ref{restricted Chern class}) 
(or possibly of (\ref{anti-matter})) must be wrong.
Clearly all the computations above were done only for the case where the curve
$A_{\La^2 V}$ is as 'good' as possible; now, the assumption of 
[\ref{DHOR}] was that in the {\em generic} case 
(though not in one or another special case which may be of phenomenological interest)
this curve is as 'good' as one could wish, namely smooth, reduced and irreducible, {\em and} that
the map $\pi: \widetilde{A}_{\La^2 V} \ra A_{\La^2 V}$ is an ordinary ramified double covering.

This assumption turns out to be wrong, however: 
the curve and the covering in question are 'special' already generically.
We will now show that already generically the curve $A_{\La^2 V}$ is
non-reduced for $n=4$ and the covering has points with special behaviour 
(this will be specified later) for $n=5$.
(For the following recall that in the group law on the elliptic fibre 
the $n$ points of $C$ on a fibre add up to zero: $\sum_{i=1}^n p_i=0$.)

\subsubsection{The case $n=4$}

For $n=4$ one has the cohomology classes 
$\widetilde{A}_{\La^2 V}=(4\si+\eta)(\eta-3c_1)$ and $A_{\La^2 V}=2(\eta-3c_1)$
and sees that if we are on a $C_V$-fibre over a point of $A_{\La^2 V}\subset B$
where two points $p_1$ and $p_2$ are $\tau$-conjugate to each other, in other 
words where $p_1+p_2=0$, then one will also have $p_3+p_4=0$,
i.e.~each point of the curve 'counts twice'. 
The curve turns out to be two times its reduced version:
the cohomology class of $A_{\La^2 V}$ shows already that this is indeed possible and
consideration [\ref{FMW}] of the $\tau$-action in coordinates shows
that the reduced curve is actually given by $a_3=0$ (cf.~footn.~\ref{class notation})
which has just the correct class $\eta-3c_1$. 
(The locus of 'special' points (which can cause deviations) 
- here the curve as a whole - 
is insofar in harmony with the correction $\Delta_4=(\eta-3c_1)c_1$ we found
as the latter is 'proportional' to (the cohomology class of) the former: 
$\Delta_4=\frac{1}{2}A_{\La^2 V}c_1$; but one can not 'turn off' the special locus
by the choice $\eta=3c_1$ which violates the effectivity of $e_4$).

\subsubsection{The case $n=5$}

For $n=5$, where one has $A_{\La^2 V}=3\eta-10c_1$, one gets similarly 
in a $C_V$-fibre over a point of our curve $A_{\La^2 V}\subset B$ that $p_3+p_4+p_5=0$; 
so here the 'doubly counting' points\footnote{where a second pair of points $\{p_k, p_l \}$ 
(where $k\neq l$) with $p_k+p_l=0$ exists 
besides the pair $\{ p_1, p_2 \}$ (so one needs to demand here $\{p_k, p_l \}\neq \{ p_1, p_2 \}$),
cf.~remark in the next paragraph}
in the base curve $A_{\La^2 V}$ (of the covering $\pi: \widetilde{A}_{\La^2 V}\ra A_{\La^2 V}$)
are just the points where, say,
$p_5$ is zero. This leads to an investigation of the intersection locus consisting of
the points where our curve meets the other curve\footnote{where one of the fibre points of $C_V$ 
is zero; it has class $\eta-5c_1$ 
which is the class of the vanishing locus of $a_5$ (cf.~footn.~\ref{class notation}), 
i.e.~of the locus of $A_V=C_V\cap \si$} 
$A_V$.
The number of these points is $\nu_5=(3\eta - 10 c_1)\cdot (\eta-5c_1)$.
The locus $A_{\La^2 V}\cap A_V$ we investigate here (as the possible source of the deviation) 
is not yet precisely the source of the correction $\Delta_5$ as the latter is not 
proportional\footnote{one can discretely tune to zero $\nu_5$ 
(by a suitable choice of $\eta-5c_1$, say $15b+6f$ on
${\bf F_1}$) while having nevertheless still $\Delta_5\neq 0$
(actually $-9$ in the mentioned example)} 
to $\nu_5$. This discrepancy is easily explained, we prefer, however, to delay this consideration
to a later section and will give first a systematic study of all relevant special loci in the 
problem in general.

We remark that the proper interpretation of the phrase 'doubly counting'
used above will also be investigated further below in sect.~\ref{resolution}: 
for the moment 'doubly counting' refers just
to points $b\in B$ which have points $p_i, p_j, p_k, p_l$ in the fibre $C_b:=\pi^{-1}(b)$ of $C_V$
over $b$ with $p_i+p_j=0$ ($i\neq j$)and $p_k+p_l=0$ ($k\neq l$) 
where the index sets $\{ i,j\}$ and $\{k,l \}$ are different: $\{i,j\}\neq \{k,l\}$. 
It will be important to distinguish the latter
condition from the stronger requirement that one has even $\{i,j\}\cap \{k,l\}=\emptyset$.
That the latter, stronger condition is also fulfilled was clear in the case of $n=4$
when the existence of a specific second pair was derived; to have the same also in the argument 
we used for $n=5$ one has to make sure
that the point playing the role of $p_5$ above was not already among the pair consisting of
the points $p_1$ and $p_2$ (which led us to a generic point of the curve $A_{\La^2 V}$ in the 
first instance); the latter possibility 
(that $p_5$ is either $p_1$ or $p_2$, where then in fact $p_1=0=p_2$) -  and its impact
of splitting different effects in the geometry which contribute to the correction - 
will be considered in detail below in sect.~\ref{resolution}. 

So we have found that the relevant curve $A_{\La^2 V}$ has special loci (even generically) 
for $n=4$ (where the curve is nonreduced) 
and $n=5$ (where 'doubly counting' points exist whose precise interpretation has yet to be
investigated); one should find that, 
if one takes into account appropriately in all computations
the special nature of the curve $A_{\La^2 V}$ 
and the special nature of the covering near the special points, the correction term vanishes.

\subsubsection{Necessity to consider also the more general case $n\geq 6$}

The correction term $\Delta_n$ exists also, however, for $n\geq 6$. Apart from the arguments given
above for the cases $n=4$ and $5$ 
one should, for reasons of mathematical consistency, therefore also find points of $A_{\La^2 V}$ 
which are 'special' for the covering $\pi: \widetilde{A}_{\La^2 V}\ra A_{\La^2 V}$ for $n\geq 6$ 
(and one should find furthermore that, 
taking these properly into account in all computations, the correction in the end vanishes). 
These considerations show that the arguments used so far are in any case incomplete. 
For the reasoning we have used up to now is no longer sufficient to treat these cases: 
in a fibre over a point of our curve $A_{\La^2 V}$ we find only that 
$p_3+p_4 + \dots + p_{n-1}+p_n=0$ and it is not immediately clear whether a further pair
$\{ p_k, p_l \}$ (where $k\neq l$) with $p_k+p_l=0$ (and $\{ 1,2 \} \cap \{ k,l \}=\emptyset$)
is enforced.

These preliminary considerations will be refined for all $n\geq 6$ in sect.~\ref{resolution}.
There we will also carefully take into account
the difference of having only $\{ 1,2 \} \neq \{ k,l \}$ or the stronger
$\{ 1,2 \} \cap \{ k,l \}=\emptyset$. Furthermore the relation to the branching phenomenon,
where in a pair $p_i, p_j$ (where $i\neq j$) with $p_i+p_j=0$ one has actually $p_i=p_j$,
will be considered. Thereby we will be able to solve the reduced problem.
After having described this we will go back in the final subsection \ref{cases 4 and 5 finally} 
again to the cases $n=4$ and $n=5$.

\section{\label{resolution}The resolution}

\resetcounter

To resolve the discrepancies above let us have now a closer look at the fibration 
structure of the ramified two-fold covering $\pi: \widetilde{A}_{\La^2 V}\ra A_{\La^2 V}$.

\subsection{\label{branching subsection}Consideration of the ramification locus}

Above the branching of the map $\pi: \widetilde{A}_{\La^2 V}\ra A_{\La^2 V}$
was computed as follows (we denote, following [\ref{DHOR}], the ramification divisor above by $R$ 
and the branching divisor below by $Br$) 
\beqa
\mbox{deg}\, Br \; = \; \mbox{deg}\, R&=& \widetilde{A}_{\La^2 V} \cdot \si_{II}\\
&=&4 \eta^2 - 2(n+4)\eta c_1 + n (n-1) c_1^2
\eeqa
The ramification locus $R$ (above) is the locus where two generically different 
(this is the definition of $\widetilde{A}_{\La^2 V}$)
fibre points $p_i, p_j$ ($i\neq j$) with $p_i=\tau p_j$
(let us call such a pair of fibre points henceforth a mirror pair for short)
come together. 
Clearly the limit point (the coalescing point) 
of such a mirror pair is a self-mirror point, i.e.~a
$2$-torsion point, i.e.~an element of $C\cap \si_{II}$; 
so one has (here the first inclusion is an identity\footnote{the limit point 
of a coalescing mirror pair is a self-mirror point;
conversely, a self-mirror point $p_*$ on a branch
of the curve (of mirror pairs) $\widetilde{A}_{\La^2 V}$ is a point where the 
two branches of $\widetilde{A}_{\La^2 V}$ meet, as the mirror partner of $p_*$ is again $p_*$ 
(such a point could also be a double point if $\widetilde{A}_{\La^2 V}$ would not be smooth)})
\beqa
R& \subset &\widetilde{A}_{\La^2 V}\cap \si_{II}\;\subset \; C_V\cap \si_{II}
\eeqa 
Points of the set $C_V\cap \si_{II}$ which are such limits, i.e.~elements of $R$, 
are in particular doubly counting points of $C_V$ 
(but this is not sufficient); so one has 
\beqa
R&\subset &r_V\cap \si_{II}\;\subset \; C_V\cap \si_{II}
\eeqa
Here the first inclusion is however not an equality: 
although a two-torsion point $p_*$ (in a fibre $C_b$ of $C_V$)
which is actually also doubly counting (in $C_b$) 
is the limit of different fibre points in neighbouring fibers 
which approach the fibre $C_b$ (containing $p_*$)
these neighbouring point pairs do not have to be mirror pairs.
Therefore one expects $\mbox{deg}\, R < r_V\cdot \si_{II}$.

Actually one has
\beqa
r_V\cdot \si_{II}&=&\widetilde{A}_{\La^2 V}\cdot \si_{II} +12 \eta c_1
\eeqa
and thus one finds indeed just the sort of relation one expects here
(we mean the relation 
$\mbox{deg}\, R =\widetilde{A}_{\La^2 V} \cdot \si_{II}< r_V \cdot \si_{II}$)
because we assume that $c_1$ is ample so that
$\eta c_1>0$ (as $\eta-n c_1$ has to be effective).
More precisely, one finds that the $\si_2$-term of $\si_{II}=\si+\si_2$
causes the difference as one has (a relation which itself is not a priori 
obvious\footnote{as $\widetilde{A}_{\La^2 V}\cdot \si $
concerns the case where the coalescing point of a {\em mirror} pair
is zero whereas in $r_V \cdot \si$ the coalescing point of a pair 
(an arbitrary pair of fibre points coming together in a special fibre) is zero.})
\beqa
\label{identity}
\widetilde{A}_{\La^2 V}\cdot \si &=&A_V \cdot A_V' \; = \; r_V \cdot \si
\eeqa
where we use the notation $A_V':=\eta-(n-1)c_1$ (as cohomology class),
a class 'derived' from the class $A_V=\eta-nc_1$.

{\em Remark:} The considerations above show also that inside the curve 
$A_V$, which consists of points of $b\in B$ where a point $p_i=0$ in the 
fibre $C_b$ of $C_V$ exists, there exist a number of $r_V\cdot \si= A_V \cdot 
A_V'$ points where actually two points of $p_i=0=p_j$ ($i\neq j$) exist in $C_b$.
Below in (\ref{first set}) we will
extrapolate this to $A_{\La^2 V}$ and will give a detailed interpretation.

One finds now for (the class of) the ramification locus $r_{\La^2 V}$ of $\pi: C_{\La^2 V}\ra B$
\beqa
r_{\La^2 V}&=&n(n-1)\Bigg[ (n-2)\eta-\frac{n^2-n-2}{4}c_1 \Bigg]\si
+(n-2)\eta \Big[ (n-2)\eta + c_1\Big]
\eeqa
One has, in analogy to (\ref{identity}), here the following relation
(again with $A_{\La^2V}':=A_{\La^2V}+c_1$)
\beqa
\label{La2 V identity}
r_{\La^2 V}\cdot \si &=&A_{\La^2V}\cdot A_{\La^2V}'
\eeqa

Let us distinguish in a more detailed manner various subsets which are
relevant in the ramification of the covering $\pi: \widetilde{A}_{\La^2 V}\ra A_{\La^2 V}$
(for the meaning of the indices $k,j$ cf.~below)
\beqa
\label{first inclusion}
\pi(\widetilde{A}_{\La^2 V}\cap \si_{II})\;\; \subset \;\;
\pi(\widetilde{A}_{\La^2 V}\cap r_V)
\;\;\;\;\;\;\;\;\;\;\;\;\;\;\;\;\;\;\;\;\;\;\;\;\;\;\;\;\;\;\;\;\;\;\;\;\;\;\;\;\;\;\;\;\;\;
(k=j \; \mbox{allowed})\;\;\\
\label{second inclusion}
\Big( \pi(\widetilde{A}_{\La^2 V}\cap r_V)\; \setminus \;
\pi(\widetilde{A}_{\La^2 V}\cap \si_{II}) \Big) \;\; \subset \;\;
r_{\La^2 V}\cap \si \;\; \subset \;\; A_{\La^2V}
\;\;\;\;\;\;\;\;\;\;\;\;\;\;\;\;\;\;\;\;\;\;\;\;\;
(k\neq j)\;\;\;
\eeqa
where the different subsets have the following interpretation
(with $p_i\in C_b:=C_V\cap \pi^{-1}(b)$)
\beqa
\label{first set}
r_{\La^2 V}\cap \si \!\!\!\! & = & \!\!\!\! \{ b\in B \; | \; p_i+p_j=0=p_k+p_l 
\;\; \mbox{where} \;\; i\neq j, k\neq l \, \mbox{and} \, \{i,j\}\neq \{k,l\}  \}\;\;\;\;\;\;\;\;\\
\label{second set}
\pi(\widetilde{A}_{\La^2 V}\cap r_V)\!\!\!\! &=&\!\!\!\! \{ b\in B \; | \; p_i+p_j=0=p_k+p_j
\;\; \mbox{where} \;\; i\neq j \; \mbox{and} \; k\neq i \; \}
\nonumber\\
\!\!\!\! &=&\!\!\!\! \{ b\in B \; | \; p_i+p_j=0 \; \mbox{and} \; p_i=p_k
\;\; \mbox{where} \;\; i\neq j \; \mbox{and} \; k\neq i \; \}\;\; \\
\label{third set}
\pi(\widetilde{A}_{\La^2 V}\cap \si_{II})\!\!\!\! &=&
\!\!\!\! \{ b\in B \; | \; p_i+p_j=0 \; \mbox{and} \;  p_i=p_j
\;\; \mbox{where} \;\; i\neq j \; \}
\eeqa
where in (\ref{second set}), (\ref{third set}) the mirror pair $\{p_i, p_j\}$ 
is assumed to occur in a family of mirror pairs consisting generically of 
different\footnote{the feature of the curve $\widetilde{A}_{\La^2 V}$ 
distinguishing it from the other components $\si|_C$ and $\si_2|_C$ of $\tau C|_C$} 
points.
In (\ref{second set}) the interpretation of $r_V$ is
that one has $k\neq i$ although it is still possible that $k=j$;
the latter case leads to the first inclusion (\ref{first inclusion})
(the condition in (\ref{third set}) describes the
ramification points of the covering $\pi: \widetilde{A}_{\La^2 V}\ra A_{\La^2V}$).
By contrast the case $k\neq j$ leads to the second inclusion 
(\ref{second inclusion}).

Let us also give the cardinalities of the sets 
(\ref{first set}) - (\ref{third set})
\beqa
r_{\La^2 V}\cdot \si & = & 
\Big[n^2-4n+4\Big]\eta^2-\Big[n^3-3n^2+n+2\Big]\eta c_1
+n(n-1) \frac{n^2-n-2}{4} \, c_1^2\;\;\;\;\;\;\; \\
\widetilde{A}_{\La^2 V}\cdot r_V&=&
\Big[3n-4\Big]\eta^2-\Big[3n^2-4n+4\Big]\eta c_1+ n(n-1)\Big[n-1\Big] c_1^2\\
\label{third cardinality}
\widetilde{A}_{\La^2 V}\cdot \si_{II}&=&
4\eta^2-\Big[2n+8\Big]\eta c_1 + n(n-1) c_1^2
\eeqa
(as on points sets $\pi$ is generically one-to-one the cardinalities can be computed 
upstairs)\footnote{we will learn later, in sect.~\ref{n=4 again}, that for this statement to be true
one has to demand $n>4$}.

In view of the inclusions (\ref{first inclusion}), (\ref{second inclusion})
also the following sets are interesting
(with the same assumption for the occurring mirror pair $\{ p_i, p_j\}$ 
as above)
\beqa
\pi(\widetilde{A}_{\La^2 V}\cap r_V) \setminus 
\pi(\widetilde{A}_{\La^2 V}\cap \si_{II}) &=&
\{ b\in B \; | \; p_i+p_j=0 \; \mbox{and} \; p_i=p_k \nonumber\\
&&\;\;\;\;\;\;\;\;\;\;\;\;
\; \mbox{where} \; i\neq j \; \mbox{and} \; k\neq i,j  \}\;\;\;\;\;\;\;\;\;
\\
\label{relevant set}
(r_{\La^2 V}\cap \si) \setminus \Big( \pi(\widetilde{A}_{\La^2 V}\cap r_V) 
\setminus \pi(\widetilde{A}_{\La^2 V}\cap \si_{II}) \Big)&=&
\{ b\in B \; | \; p_i+p_j=0=p_k+p_l \nonumber\\
&& 
\;\;\;\;\;\;\;\;\;\;\;\;\;
\mbox{where} \ i\neq j, k\neq l \nonumber\\
&&
\;\;\;\;\;\;\;\;\;\;\;\;\;
\, \mbox{and}\, \{i,j\}\cap \{k,l\}=\emptyset  \} 
\;\;\;\;\;\;\;\;\;
\eeqa
(i.e.~in both cases all indices are different).
Let us also introduce their cardinalities
\beqa
\epsilon &:=& \# \Big[ \pi(\widetilde{A}_{\La^2 V}\cap r_V) \setminus 
\pi(\widetilde{A}_{\La^2 V}\cap \si_{II}) \Big]\\
\label{relevant cardinality definition}
\delta &:= & \# \Bigg[ (r_{\La^2 V}\cap \si) \setminus 
\Big( \pi(\widetilde{A}_{\La^2 V}\cap r_V) \setminus 
\pi(\widetilde{A}_{\La^2 V}\cap \si_{II}) \Big) \Bigg]
\eeqa
$\epsilon$ and $\delta$ cover the cases 
$\{i,j\}\cap \{k,l\}\neq \emptyset$ and $=\emptyset$, resp.~(always $\{i,j\}\neq \{k,l\}$).
One has
\beqa
\epsilon &=& 
\widetilde{A}_{\La^2 V}\cdot r_V - \widetilde{A}_{\La^2 V}\cdot \si_{II}
\nonumber\\
&=&\Big[3n-8\Big]\eta^2-\Big[3n^2-6n-4\Big]\eta c_1 + n(n-1) \Big[n-2\Big] c_1^2
\\
\label{relevant cardinality}
\delta &=& 
r_{\La^2 V}\cdot \si - 
\Big( \widetilde{A}_{\La^2 V}\cdot r_V  - \widetilde{A}_{\La^2 V}\cdot \si_{II}\Big)
\nonumber\\
&=&\Big[n^2-7n+12\Big]\eta^2-\Big[n^3-6n^2+7n+6\Big]\eta c_1 
+ n(n-1) \frac{n^2-5n+6}{4}c_1^2\;\;\;\;\;\;\;\;
\eeqa
From the interpretation (\ref{relevant set}) one learns that the set of points
in the curve $A_{\La^2 V}$ whose cardinality $\delta$ is computed 
in (\ref{relevant cardinality}) is the set of points where the generically
two-fold covering $\widetilde{A}_{\La^2 V} \ra A_{\La^2 V}$ is actually
a four-fold one, consisting of the points $p_i, p_j, p_k, p_l$:~these points are all distinct
because $p_i$ and $p_j$ (and similarly $p_k$ and $p_l$) are distinct 
(because the point locus $\pi(\widetilde{A}_{\La^2 V} \cap \si_{II})$ over 
which they can coincide is generically distinct from the one in question here)
and none of the points $p_k, p_l$ can equal any of the other points
$p_i, p_j$ because the set 
$(\widetilde{A}_{\La^2 V}\cap r_V)\setminus (\widetilde{A}_{\La^2 V}\cap \si_{II})$ was taken out
(all indices here are different). 

Here one has to make sure that these second special pairs $\{p_k, p_l\}$ with $p_k+p_l=0$ 
(besides the first, ordinary pair $\{p_i, p_j\}$ with $p_i+p_j=0$)
above in $C_V$
are actually elements of the {\em curve} $\widetilde{A}_{\La^2 V}$; this means that they too are
part of a continuous family of such mirror pairs and not just isolated point pairs with the
mirror property. For note that $\widetilde{A}_{\La^2 V}$ was {\em defined} as a divisor by 
\beqa
\tau C_V|_{C_V}&=&\si_{II}|_{C_V}+\widetilde{A}_{\La^2 V}
\eeqa
(i.e.~essentially as a {\em curve}).
It was {\em not defined}, (a priori) differing, as the {\em set} given by 
\beqa
\tau C_V \cap C_V &\stackrel{?}{=}& (\si_{II}\cap C_V) \cup \widetilde{A}_{\La^2 V}
\eeqa 
When one just considers the covering $\pi: \widetilde{A}_{\La^2 V}
\ra A_{\La^2 V}$ one learns only that, going away continuously in $A_{\La^2 V}$ from a special
point, at least one of the two mirror pairs $\{p_i, p_j\}$ and $\{p_k, p_l\}$ above must have
also a continuous continuation which then constitutes a local part of the covering curve; the
other mirror pair might, a priori, be an isolated pair. 

The existence of the second pair was implied, however, by consideration of the set
$r_{\La^2 V}\cap \si$. The ramification curve $r_{\La^2 V}$ is the locus 
where two local surface branches of the covering 
$\pi_{\La^2 V}: C_{\La^2 V}\ra B$ meet. These local surface branches meet also in $X$ the embedded
surface $B$, i.e.~$\si(B)$.\footnote{we usually identify the base $B$, the section $\si:B\ra X$, 
its image in $X$ and its cohomology class}
These intersections constitute local parts of the curve $A_{\La^2 V}=C_{\La^2 V}\cap \si$.
These local parts give, here in the base curve $A_{\La^2 V}$, 
the respective continuations (away from the special point in $r_{\La^2 V}\cap \si$) 
which correspond to the searched for continuations of the two mirror pairs above.

One still has to clarify whether these special points in the base curve $A_{\La^2 V}$ of the 
fibration are singularities (double points) or smooth points like 
ramification points\footnote{which in a dimensionally reduced, real picture 
can not be distinguished from double points: recall the standard picture
for a representation of an elliptic curve as fourfold ramified double cover of ${\bf P^1}$}
(in some {\em other} covering) where different branches meet. According to [\ref{DHOR}], sect.~7.5, 
the generic curve $A_{\La^2 V}$ is smooth, so it has to be the latter option.
This property of a {\em smooth} meeting of local branches of $A_{\La^2 V}=C_{\La^2 V}\cap \si$
in a special point can be considered (for yet another heuristic argument cf.~below)
as being inherited from the smooth meeting of local branches of $C_{\La^2 V}$
in the ramification curve $r_{\La^2 V}$ 
(which generically also is not a curve of double points).

Let us recapitulate how the described non-generic behaviour of the fibre 
of the map $\widetilde{A}_{\La^2 V} \ra A_{\La^2 V}$ is possible.
Note that when at such a special fibre the curve 
$\widetilde{A}_{\La^2 V}\subset C_V$ upstairs is projected down to the curve
$A_{\La^2 V}=C_{\La^2 V}\cap B \subset B$ then two local branches of $A_{\La^2 V}$ meet there
(projections of local parts of $\widetilde{A}_{\La^2 V}$ 
representing continuations of the mirror pairs $\{p_i, p_j\}$ and $\{p_k, p_l\}$). 
However, $A_{\La^2 V}$ does not have double points there: the generic
$A_{\La^2 V}$ is smooth [\ref{DHOR}]. We now give a second heuristic explanation of the latter fact.

One may compare the case $A_V=C_V \cap B$ which also
does {\em not} have double points at the set of special points $r_V\cap \si$
(which are also 'doubly counting' as one has there fibre points with $p_i=0=p_j$
($i\neq j$)).
The surface $C_V$ is completely 'general', i.e.~it is just a generic member of the
{\em full} linear system $|C_V|$, and actually $A_V=C_V\cap B$ is also completely generic
in its linear system $|A_V|$ in $B$ and so the generic member is smooth; 
at the points of $r_V \cap \si$ two local branches of
$A_V$ meet in a {\em non-singular} point like in a ramification point. 
For $C_{\La^2 V}$ the situation is  different at first: this surface 
is not a generic member of its linear system, rather it arises fibrewise
by adding up (in the group law on the fibre) the corresponding fibre points
of $C_V$ (which themselves were generic); however, when now intersecting with $\si$
(i.e.~$\si(B)$, the embedded base surface), one finds that $A_{\La^2 V}$ is nevertheless 
still a generic member of its own linear system in $B$ and so nonsingular.

As this point is relevant\footnote{\label{cancel out}in the case of double points 
of $A_{\La^2 V}$ the resulting correction effect would cancel out, 
cf.~sect.~\ref{impact subsection}} let us consider this in detail. 
Assume we deform $A_{\La^2 V}$ to a 
curve\footnote{still in the same cohomology class $(n-2)\eta-\frac{n(n-1)}{2}c_1$ 
in $H^2(B, {\bf Z})$}
$\ov{A}_{\La^2 V}$:
can we find a deformed surface\footnote{still in the cohomology class
$n\si+\eta$ in $H^2(X, {\bf Z})$ (and with fibre points adding to zero)} $\ov{C}_V$ 
with $\ov{A}_{\La^2 V}=\ov{C}_{\La^2 V}\cap \si$ where
$\ov{C}_{\La^2 V}$ is derived in the usual way from $\ov{C}_V$ (that is by building fibrewise
the $p_i+p_j$ for $i< j$)? If true, the generic curve of class $(n-2)\eta-\frac{n(n-1)}{2}c_1$ 
arises indeed as an $A_{\La^2 V}=C_{\La^2 V}\cap \si$ from a generic $C_V$.
The question has two parts: 1.) whether $\ov{A}_{\La^2 V}$ can be understood
as $S\cap \si$ for a surface $S$ of class $\frac{n(n-1)}{2}\si +(n-2)\eta$, 
and 2.) whether $S$ is a $\ov{C}_{\La^2 V}$.

Concerning 1.)~note that among the moduli of possible surfaces $S$
(essentially the parameters in the corresponding equation) there is a subset $N$ of parameters
which arises when one considers the ensuing equation for $S\cap \si$; there is no obstruction
to complete {\em arbitrary} parameters of $N$ (for the equation of the curve $S\cap \si$) 
to a complete set of parameters for (an equation of) $S$.
For 2.)~note that the parameter 
restriction\footnote{that the $\frac{n(n-1)}{2}$ fibre points $q_k$ of $S$  
arise as $p_i+p_j$ (with $i,j=1, \dots, n$ and $i< j$) where $\sum p_i=0$} 
which assures that a general $S$ is a $\ov{C}_{\La^2 V}$ concerns only 
'vertical' parameters whereas $N$ refers to 'horizontal' parameters 
(suggesting that the latter remain unrestricted and generic).

\subsection{\label{impact subsection}The impact of the special points}

For a proper {\em bookkeeping} of the effect of the special points
let us define {\em formally} a corrected Euler number of the curve $A_{\La^2 V}$
which includes besides the ordinary expression from adjunction 
(for a generic smooth curve of the class of $A_{\La^2 V}$)
also a correction term
\beqa
\label{correction term}
e_{A_{\La^2 V}}^{corr}&=&e_{A_{\La^2 V}}^{ord}-2\Delta_n
\eeqa
With the inclusion of this correction term
both expressions in question now come out correctly:
on the one hand the Euler number computation for the curve
$\widetilde{A}_{\La^2 V}$ from the ramified covering
$\pi: \widetilde{A}_{\La^2 V}\ra A_{\La^2 V}$
\beqa
\label{corrected version of double covering formula}
e_{\widetilde{A}_{\La^2 V}}&=&2e_{A_{\La^2 V}}^{corr}-\mbox{deg} \, Br
\eeqa
and the generation number computation from the net amount of matter
on the other
\beqa
\label{generation number computation from the net amount of matter}
h^1(\La^2 V)-h^1(\La^2 \ov{V})&=&\mbox{deg}(L_{\La^2 V}\otimes K_B)|_{A_{\La^2 V}}
+\frac{1}{2}e_{A_{\La^2 V}}^{corr}
\eeqa

Let us emphasize that the actual motivation of the correction in the Euler number
came from the non-standard behaviour of the covering
$\pi: \widetilde{A}_{\La^2 V}\ra A_{\La^2 V}$ at the special points. 
Taking into account this non-standard behaviour 
one has of course to go similarly now through the matter computation of the generation number 
from the beginning and to find that the final formula {\em turns out} again to be the previous
original formula (for the nonsingular case), now just with the 'renormalized' corrected 
Euler number (instead of the ordinary one); that actually all the effects of the special 
points for {\em this} computation can be absorbed just in a redefined Euler number would 
deserve a detailed reasoning. What we have shown is that this is so {\em effectively}.

Having seen that the redefinition (\ref{correction term}) works
let us now explain why it should do so.
The crucial identification of the correction term $\Delta_n$ is expressed in the relation
(cf.~(\ref{relevant cardinality definition}))
\beqa
\fbox{\mbox{$-2\Delta_n\; = \; \delta \; = \; r_{\La^2 V}\cdot \si - 
\Big( \widetilde{A}_{\La^2 V}\cdot r_V 
- \widetilde{A}_{\La^2 V}\cdot \si_{II}\Big)$}}
\eeqa
From the interpretation in (\ref{first set}) and (\ref{relevant cardinality})
one finds that this non-negative number is the number of special points of 
$A_{\La^2 V}$ with four rather than only the generic two preimages in
the curve $\widetilde{A}_{\La^2 V}$ upstairs. The reason for the 
(positive) occurrence of $\delta$ in the {\em effective correction} (\ref{correction term}), 
which then occurs in the correct double covering formula 
(\ref{corrected version of double covering formula}),
is just that over these special points of the curve $A_{\La^2 V}$ downstairs lie
not two but four points in the curve $\widetilde{A}_{\La^2 V}$ upstairs. Or, more formally:
if the mentioned locus of points is taken out on the left hand side 
of the double covering formula with this fourfold multiplicity 
and on the right hand side with simple multiplicity (as correction to
the Euler number, i.e.~inside the bracket which is multiplied by the
covering degree)
one arrives {\em effectively} indeed at the corrected version of the formula
as given in (\ref{corrected version of double covering formula})
\beqa
e_{\widetilde{A}_{\La^2 V}}-4\delta&=&2(e_{A_{\La^2 V}}^{ord}-\delta)
-\mbox{deg} \, Br\\
\label{standard interpretation}
\Longrightarrow e_{\widetilde{A}_{\La^2 V}}&=&2(e_{A_{\La^2 V}}^{ord}+\delta)
-\mbox{deg} \, Br
\eeqa
We emphasize again (cf.~footn.~\ref{cancel out}) that it is decisive that the special points are, 
despite first appearance perhaps, not double points 
(but rather ramification points of local branches of the curve $A_{\La^2 V}$ in the base).
If they would have been double points one would have had 
$e_{A_{\La^2 V}}^{ord}=e_{A_{\La^2 V}}^{standard}-\delta$ and the whole effect would have
cancelled out (here 'standard' refers to the standard formula from the adjunction computation
which only for a smooth curve gives the ordinary Euler number).

This reasoning explains the correction in the double
covering formula (\ref{corrected version of double covering formula})
which compares the Euler numbers of the curves
$\widetilde{A}_{\La^2 V}$ upstairs and $A_{\La^2 V}$ downstairs (the auxiliary problem).
We recall that, by contrast, the fact that just this effective correction occurs also again
in the computation (\ref{generation number computation from the net amount of matter}) 
of the generation number from the net amount of matter (the original problem),
would deserve a second start and going through that computation
while taking into account the influences of the mentioned special points.
We stress that we do {\em not} claim that the correct matter computation 
{\em arises} from the use of a corrected Euler number (as causation); 
rather one has also to review (\ref{restricted Chern class}).
Our point is that the complicated inner structure of the correction term, 
which was completely explained in the auxiliary problem, suggests that this explanation 
(the uncovered geometric subtlety)
captures also the reason for the identical correction in the original problem.

{\em Remark:} One might phrase the whole phenomenon by starting from the algebraic-geometric 
object (more general than a curve) given by the curve $A_{\La^2 V}$ together
with some of its points having higher (double) multiplicity; 
{\em effectively} then the results (\ref{corrected version of double covering formula}) 
and (\ref{generation number computation from the net amount of matter}) 
work out correctly just with the Euler number contribution
$e_{A_{\La^2 V}}^{for}+\delta$ of this more subtle object.
That this second effect in (\ref{generation number computation from the net amount of matter}) 
also takes place appropriately (as demanded by the result of the Chern class computation) 
is now however clearly suggested.

\subsection{\label{cases 4 and 5 finally}A second look at the cases $n=4$ and $n=5$}

Having understood how even\footnote{where (in contrast to the cases $n=4$ or $5$) 
our previous reasoning just from the relation $\sum_{i=3}^n p_i=0$ 
in a fibre over a point of our curve $A_{\La^2 V}$ was no longer sufficient to capture
the special points} 
in the cases $n\geq 6$ special points arise which can explain the deviations from the
expected expressions in the mentioned formulae, let us now come back and have a second look
at the special, but physically important cases $n=4$ and $5$. 

\subsubsection{\label{n=4 again}The case $n=4$ again}

In the case $n=4$ the general fibre relation $p_1+p_2+p_3+p_4=0$ in $C_V$ shows immediately
the nonreduced character of the curve $A_{\La^2 V}$ as we described earlier in 
sect.~\ref{cases 4 and 5}. This will, in this special case, modify of course 
the considerations done for the general case $n\geq 6$. 
We have two points to tackle: first the discrepancy between the generally computed number $\delta$ of 
special points (on the one hand) and the true nature of the special locus as being given here by
the curve $A_{\La^2 V}$ as a whole (on the other hand); and secondly
how the appropriate relation between the Euler numbers of $\widetilde{A}_{\La^2 V}$ 
and $A_{\La^2 V}$ is constructed, 
especially given the two facts of $A_{\La^2 V}$ being non-reduced and
the special locus consisting of the base curve $A_{\La^2 V}$ as a whole
(instead of being given by a finite set of special points).

Let us start with the first point, the proper adjustment of $\delta$. We had found 
a proportionality between $\delta_4=-2\Delta_4=-2(\eta-3c_1)c_1$ and the 'special' locus, 
here actually the curve $A_{\La^2 V}$ (of cohomology class $2(\eta-3c_1)$) as a whole.
Let us now 
point out where the general argument (for $n\geq 6$) fails that $\delta$ gives the cardinality of
the locus of special points. One finds in this case that the ramification curve $r_{\La^2 V}$
of the covering $\pi_{\La^2 V}: C_{\La^2 V}\ra B$ is actually reducible
(with one component being furthermore non-reduced as $A_{\La^2 V}=2A_{\La^2 V}^{red}$):
\beqa
r_{\La^2 V}&=&r_{\La^2 V}^{(\delta)}+r_{\La^2 V}^{(\epsilon)}\nonumber\\
           &=&\si_{II}|_{C_{\La^2 V}}+r_{\La^2 V}^{(\epsilon)}\nonumber\\
           &=&A_{\La^2 V}\si + \si_2|_{C_{\La^2 V}}+r_{\La^2 V}^{(\epsilon)}
\eeqa
We can here distinguish a $\delta$-component $r_{\La^2 V}^{(\delta)}$ and an $\epsilon$-component
$r_{\La^2 V}^{(\epsilon)}$ of $r_{\La^2 V}$, being given by the cases of 
$p_i+p_j=p_k+p_l$ (where $i\neq j$ and $k\neq l$, and of course $\{i, j\}\neq \{k, l\}$) 
with either $\{i, j\}\cap \{k, l\}=\emptyset$ or  $\{i, j\}\cap \{k, l\}\neq\emptyset$, 
respectively (cf.~an analogous remark after (\ref{relevant cardinality definition})):
in the first case one finds with $p_1+p_2=p_3+p_4$ that $p_1+p_2\in \si_{II}|_{C_{\La^2 V}}$
(because one also has $p_1+p_2=-(p_3+p_4)$), which itself has the two components 
$\si|_{C_{\La^2 V}}=A_{\La^2 V}\si$ and $\si_2|_{C_{\La^2 V}}$; in the second case one gets a
a remaining component of class $r_{\La^2 V}-r_{\La^2 V}^{(\delta)}$
\beqa
r_{\La^2 V}^{(\epsilon)}
&=&(C_{\La^2 V}+c_1)|_{C_{\La^2 V}}-\si_{II}|_{C_{\La^2 V}}\nonumber\\
&=&2\Big(\si+\eta-c_1\Big)\Big|_{C_{\La^2 V}}\nonumber\\
&=&4\Big[(4\eta-6c_1)\si+\eta^2-\eta c_1\Big]
\eeqa

Here one finds therefore, as we expected above to find in this special case $n=4$, 
for the set $\{ b\in B \, | \, p_i+p_j=0=p_k+p_l \;\mbox{where} \;
i\neq j, k\neq l \;\mbox{and} \; \{ i,j\}\cap \{ k,l\} =\emptyset \}$ (cf.~(\ref{relevant set}))
just the reduced component $A_{\La^2 V}^{red}$; as an additional numerical check
one convinces oneself that $r_{\La^2 V}^{(\epsilon)}\cdot \si =4(\eta-3c_1)(\eta-2c_1)$
such that one has indeed
\beqa
\epsilon&=& r_{\La^2 V}^{(\epsilon)}\cdot \si 
\eeqa
Actually the situation here is still somewhat more complicated, however. The proper counting
turns out in this case to employ the halved quantities instead of the general formal expressions: 
for $r_{\La^2 V}^{(\epsilon)}\cdot \si$ one has to use it halved 
because it comes with double multiplicity (like $2A_{\La^2 V}^{red}\si$), 
and similarly for $\epsilon$ the argument
given after (\ref{third cardinality}) is no longer true such that one has actually to use
$\epsilon^{true}:=\frac{1}{2}\epsilon$. After these adjustments it is still true, however, that
\beqa
\epsilon^{true}&=& \frac{1}{2} r_{\La^2 V}^{(\epsilon)}\cdot \si 
\eeqa
For $\epsilon^{true}$ the argument for the halving is that the cardinality of the subset 
$\pi(\widetilde{A}_{\La^2 V}\cap r_V)\setminus \pi(\widetilde{A}_{\La^2 V}\cap \si_{II})
=\pi(\widetilde{A}_{\La^2 V}\cap r_V \setminus \widetilde{A}_{\La^2 V}\cap \si_{II})$
of $\pi(\widetilde{A}_{\La^2 V}\cap r_V)$ is $\frac{1}{2}(\widetilde{A}_{\La^2 V}\cdot r_V
- \widetilde{A}_{\La^2 V} \cdot \si_{II})$
because the case that $p_3$ is $p_1$, say (the case of $p_2=p_1$ was excluded here),
entails that also $p_4=-p_3=-p_1=p_2$; thus one has two intersections above in 
$\widetilde{A}_{\La^2 V}\cap r_V\setminus \widetilde{A}_{\La^2 V}\cap \si_{II}$ 
over one and the same point $b\in  A_{\La^2 V}\subset B$ below.
For the same reason the other object, $r_{\La^2 V}^{(\epsilon)}\cdot \si$
comes with an inherent doubled multiplicity (relative to the set-theoretic version).

This resolves the first discrepancy in this case of $n=4$: the special locus is the base curve
as a whole and the appropriate refined consideration of $r_{\La^2 V}$ is in harmony with the proper 
$\delta$- and $\epsilon$-computations (the formal expression with $\delta_4<0$ causes no problem).

Now we have to tackle the second point: the Euler number computation for the covering
$\pi:\widetilde{A}_{\La^2 V}\ra A_{\La^2 V}$. The standard interpretation of the 
relation (\ref{standard interpretation}), which {\em as a purely numerical relation}
is still true also for $n=4$, is no longer appropriate now, of course. The proper interpretation
follows now the following computation
\beqa
e_{\widetilde{A}_{\La^2 V}}&=&
4e_{A_{\La^2 V}^{red}}-2\epsilon^{true}-\widetilde{A}_{\La^2 V}\cdot \si_{II}
\eeqa
where 
\beqa
e_{\widetilde{A}_{\La^2 V}}          &=&-4(\eta-3c_1)(3\eta-7c_1)\\
e_{A_{\La^2 V}^{red}}                &=&- (\eta-3c_1)(\eta-4c_1)\\
\epsilon^{true}                      &=& 2(\eta-3c_1)(\eta-2c_1)\\
\widetilde{A}_{\La^2 V}\cdot \si_{II}&=&4(\eta-3c_1)(\eta-c_1)
\eeqa
Here one has a more complicated ramification structure: in addition to the usual simple ramification
related to $\widetilde{A}_{\La^2 V}\cap \si_{II}$ 
(where two of the now four branches of $\widetilde{A}_{\La^2 V}$ come together) 
one has now also the double ramification related to
a number of $\epsilon^{true}$ points in the base curve $A_{\La^2 V}^{red}$ 
over which two times two of the four branches come together 
(these points also turn out to be ramification points, not double points):
this represents the coincidences from $p_3=p_1$, say, (and so also $p_4=p_2$)
which are typical for the $\epsilon$ contribution.

\subsubsection{\label{n=5 again}The case $n=5$ again}

Things are different in many respects in the case $n=5$. 
In that case we had pointed out in sect.~\ref{cases 4 and 5} that
the special locus we had found (as the possible source of the deviation) 
was not yet precisely the source of the correction $\Delta_5$ as the latter was not proportional
to $\nu_5$. This discrepancy is explained, however, easily as follows.
To have in a fibre of $C_V$ four points $p_i, p_j, p_k, p_l$ with $p_i+p_j=0$ (where $i\neq j$)
and $p_k+p_l=0$ (where $k\neq l$)
{\em with all four indices $i,j,k,l$ different}\footnote{the meaning of the cases where
$\{ i,j\} \neq \{ k,l\}$ but  $\{ i,j\} \cap \{ k,l\}\neq \emptyset$ 
was investigated in sect.~\ref{branching subsection}}
one has to make sure that the point playing the role of $p_5$ 
(in the argument given above in sect.~\ref{cases 4 and 5}) 
was not already among the pair consisting of the points $p_1$ and $p_2$
(which led us to a generic point 
of the curve $A_{\La^2 V}$ in the first instance; so here $\{ 1,2 \}=\{ i,j \}$); 
the latter possibility (of $p_5=p_1$, say, where then in fact $p_1=0=p_2$ because we imposed 
the condition $p_5=0$ in the argument of sect.~\ref{cases 4 and 5}) corresponds to the
locus $\widetilde{A}_{\La^2 V}\cap \si$ (which therefore has to be taken out) of cardinality
$\widetilde{A}_{\La^2 V}\cdot \si=\eta^2-9\eta c_1 +20 c_1^2$. 
One finds 
$\nu_5 - \widetilde{A}_{\La^2 V}\cdot \si$ as cardinality of the 'proper' special locus
and realizes furthermore that 
(for $\delta_5$ cf.~(\ref{relevant set}) and (\ref{relevant cardinality definition}))
\beqa
\delta_5&=&\nu_5 - \widetilde{A}_{\La^2 V}\cdot \si
\eeqa
restoring thus the proportionality to $\Delta_5$ (as one has, as always, $\delta_5=-2\Delta_5$).

So here, in contrast to the case $n=4$, the curve $A_{\La^2 V}$ is reduced and a proper computation,
even along the lines of sect.~\ref{cases 4 and 5} 
(avoiding the more general considerations from sect.~\ref{branching subsection}), 
gives the correct number of special points 
(this corresponds to the first point we had to tackle in the case $n=4$); 
so the interpretation of the Euler number computation 
for the covering $\pi:\widetilde{A}_{\La^2 V}\ra A_{\La^2 V}$ 
given in sect.~\ref{impact subsection} 
(whose adjustment constituted the second point we had to treat for $n=4$) 
remains completely intact here.

Needless to say, the cautionary remarks made at the end of sect.~\ref{impact subsection}
concerning an extension of the analysis of the impact of the locus of special points 
from the analogous auxiliary problem (of the Euler number computation) 
to the proper original problem (of the matter computation) 
remain in order also for the cases $n=4$ and $5$.

\section{\label{Conclusion}Conclusion}

A particularly relevant case of heterotic model building concerns supersymmetric GUT models 
with GUT group $G\! =\! SU(5)$ or $SO(10)$
(for a list containing some of the many important contributions in this field cf.~[\ref{het}]). 
These cases correspond to the structure group $H_V$
of the heterotic bundle $V$ being $SU(5)$ and $SU(4)$, resp.~$\!\!$.
Consideration of the number of matter multiplets leads to computation of the bundle cohomology of
$V$ and $\La^2 V$ (or complex conjugates); the case of the Higgses leads to
consideration of the latter.

In the framework of the spectral cover approach [\ref{FMW}] to $SU(n)$ bundles 
on an elliptic Calabi-Yau threefold $\pi:X\ra B$
one considers an $n$-fold ramified cover $C_V$ of $B$ (together with a line bundle
on $C_V$ which, in the usual construction, is essentially fixed). 
The intersection curve $A_V:=C_V\cap B$ plays an important role: 
cohomological computations of $V$ over $X$
are reduced to corresponding computations for a line bundle over $A_V$.

For the computation of the number of Higgses thus the curve $A_{\La^2 V}$ is particularly 
important\footnote{From the perspective of the original bundle $V$ itself 
this is the locus of points in $B$
where (essentially) in the corresponding fibre of $C_V$ two points ($i\neq j$) fulfil $p_i+p_j=0$.
Thus $A_{\La^2 V}$ has as double cover the non-trivial divisor component 
$\widetilde{A}_{\La^2 V}$ in $C_V\cap \tau C_V$
(the other components being given by $C_V\cap \si_{II}$).}: 
as mentioned one needs to compute
the Chern class of the relevant line bundle $L_{\La^2 V}$, 
at least when restricted to this intersection curve; 
this is formula\footnote{i.e.~(138) of [\ref{DHOR}], 
relying on (133) which assumes $\pi \!\! : \!\! \widetilde{A}_{\La^2 V}\!\ra \! A_{\La^2 V}$
to be an ordinary (ramified) covering} 
(\ref{restricted Chern class}), which uses the double cover relation
of $\widetilde{A}_{\La^2 V}$ and $A_{\La^2 V}$. We will consider just the {\em generic case} of this set-up
(whereas [\ref{DHOR}] treats in detail a {\em specific example}).

Evaluating now this computation on the matter curve $A_{\La^2 V}$ 
just for the net amount $h^1(\La^2 V)-h^1(\La^2 \ov{V})$ of the corresponding matter
one finds, cf.~(\ref{corrected bound}), an additional ('correction') term $\Delta_n$ 
relative to the expectation from the index computation from $c_3(\La^2 V)$.

As this surprising deviation is puzzling we switch to an auxiliary 
problem\footnote{where (\ref{restricted Chern class}) is not concerned 
but only the assumptions on which it itself relies, that 
$\pi \! : \widetilde{A}_{\La^2 V}\ra A_{\La^2 V}$ is an ordinary (ramified) double covering
between generically (irreducible, reduced and) smooth curves}
which involves the same double covering $\pi \!\! : \!\! \widetilde{A}_{\La^2 V}\!\ra \! A_{\La^2 V}$:
the comparison of Euler numbers of the two curves (taking into account the ramification).
We find again a deviation from the ordinarily expected relation, and again with just the 
expression $\Delta_n$ as correction.

We can explain, by a geometric subtlety 
which constitutes a deviation (details are in sect.~\ref{resolution})
from the expected standard covering picture for 
$\pi \! : \! \widetilde{A}_{\La^2 V}\!\ra \! A_{\La^2 V}$, 
the mismatch in the latter problem completely,
i.e.~we can derive just the needed correction. As the correction term itself has a 
relatively complicated parametric structure we argue that this captures already the
rationale behind the (identical) correction in the original problem.

I thank the DFG for support in the project CU 191/1-1 
and SFB 647 and the FU Berlin for hospitality.

\section*{References}
\begin{enumerate}

\item
\label{FMW}
R.~Friedman, J.~Morgan and E.~Witten, {\em Vector Bundles and F-Theory},
hep-th/9701162, Comm. Math. Phys. {\bf 187} (1997) 679.
 
\item
\label{C}
G.~Curio, {\em Chiral Matter and Transitions in Heterotic String Models},
hep-th/9803224, Phys.Lett. {\bf B435} (1998) 39.

\item
\label{DHOR}
R.~Donagi, Y.-H.~He, B.A.~Ovrut and R.~Reinbacher,
{\em The Particle Spectrum of Heterotic Compactifications},
arXiv:hep-th/0405014, JHEP {\bf 0412} (2004) 054.

\item
\label{het}
Andre Lukas, Burt A. Ovrut, Daniel Waldram,
{\em Non-standard embedding and five-branes in heterotic M-Theory},
arXiv:hep-th/9808101, Phys.Rev. D59 (1999) 106005\\
Ron Donagi, Andre Lukas, Burt A. Ovrut, Daniel Waldram,
{\em Non-Perturbative Vacua and Particle Physics in M-Theory},
arXiv:hep-th/9811168, JHEP 9905 (1999) 018\\ 
Ron Donagi, Andre Lukas, Burt A. Ovrut, Daniel Waldram,
{\em Holomorphic Vector Bundles and Non-Perturbative Vacua in M-Theory},
arXiv:hep-th/9901009, JHEP 9906:034,1999\\ 
Andre Lukas, Burt A. Ovrut, Daniel Waldram,
{\em Heterotic M-Theory Vacua with Five-Branes},
arXiv:hep-th/9903144, Fortsch.Phys.48:167-170,2000\\
Ron Donagi, Burt A. Ovrut, Daniel Waldram,
{\em Moduli Spaces of Fivebranes on Elliptic Calabi-Yau Threefolds},
arXiv:hep-th/9904054, JHEP 9911 (1999) 030\\ 
Burt A. Ovrut,
{\em N=1 Supersymmetric Vacua in Heterotic M-Theory},
arXiv:hep-th/9905115,
Lectures presented at the APCTP Third Winter School on "Duality in Fields and Strings", 
February, 1999, Cheju Island, Korea\\ 
Andre Lukas, Burt A. Ovrut,
{\em Symmetric Vacua in Heterotic M-Theory},
arXiv:hep-th/9908100\\
Ron Donagi, Burt A. Ovrut, Tony Pantev, Daniel Waldram,
{\em Standard Models from Heterotic M-theory},
arXiv:hep-th/9912208, Adv.Theor.Math.Phys. 5 (2002) 93-137\\ 
Ron Donagi, Burt A. Ovrut, Tony Pantev, Daniel Waldram,
{\em Standard Model Vacua in Heterotic M-Theory},
arXiv:hep-th/0001101,
Talk given at STRINGS'99, Potsdam, Germany, July 19-24, 1999 \\
Burt A. Ovrut, Tony Pantev, Jaemo Park,
{\em Small Instanton Transitions in Heterotic M-Theory},
arXiv:hep-th/0001133, JHEP 0005 (2000) 045\\ 
Ron Donagi, Burt Ovrut, Tony Pantev, Dan Waldram,
{\em Standard-Model Bundles on Non-Simply Connected Calabi--Yau Threefolds},
arXiv:hep-th/0008008, JHEP 0108:053,2001\\ 
Eduardo Lima, Burt Ovrut, Jaemo Park, René Reinbacher,
{\em Non-Perturbative Superpotentials from Membrane Instantons in Heterotic M-Theory},
arXiv:hep-th/0101049, Nucl.Phys. B614 (2001) 117-170\\ 
Eduardo Lima, Burt Ovrut, Jaemo Park,
{\em Five-Brane Superpotentials in Heterotic M-Theory},
arXiv:hep-th/0102046, Nucl.Phys. B626 (2002) 113-164\\ 
Burt A. Ovrut, 
{\em Lectures on Heterotic M-Theory},
arXiv:hep-th/0201032,
Lectures presented at the TASI 2000 
School on Strings, Branes and Extra Dimensions, Boulder, Co, June 3-29, 2001\\ 
Evgeny Buchbinder, Ron Donagi, Burt A. Ovrut,
{\em Vector Bundle Moduli and Small Instanton Transitions},
arXiv:hep-th/0202084, JHEP 0206 (2002) 054\\ 
Evgeny I. Buchbinder, Ron Donagi, Burt A. Ovrut,
{\em Superpotentials for Vector Bundle Moduli},
arXiv:hep-th/0205190, Nucl.Phys.B653:400-420,2003\\
Evgeny I. Buchbinder, Ron Donagi, Burt A. Ovrut,
{\em Vector Bundle Moduli Superpotentials in Heterotic Superstrings and M-Theory},
arXiv:hep-th/0206203, JHEP 0207 (2002) 066\\
Burt A. Ovrut, Tony Pantev, Rene Reinbacher,
{\em Torus-Fibered Calabi-Yau Threefolds with Non-Trivial Fundamental Group},
arXiv:hep-th/0212221, JHEP 0305 (2003) 040\\
Burt A. Ovrut, Tony Pantev, Rene Reinbacher,
{\em Invariant Homology on Standard Model Manifolds},
arXiv:hep-th/0303020, JHEP 0401 (2004) 059\\ 
Yang-Hui He, Burt A. Ovrut, Rene Reinbacher,
{\em The Moduli of Reducible Vector Bundles},
arXiv:hep-th/0306121,JHEP 0403 (2004) 043\\ 
Ron Donagi, Burt A.Ovrut, Tony Pantev, Rene Reinbacher,
{\em SU(4) Instantons on Calabi-Yau Threefolds with $Z_2$ x $Z_2$ Fundamental Group},
arXiv:hep-th/0307273, JHEP0401:022,2004\\ 
Ron Donagi, Yang-Hui He, Burt A. Ovrut, Rene Reinbacher,
{\em Moduli Dependent Spectra of Heterotic Compactifications},
arXiv:hep-th/0403291,Phys.Lett.B598:279-284,2004\\ 
Ron Donagi, Yang-Hui He, Burt A. Ovrut, Rene Reinbacher,
{\em The Particle Spectrum of Heterotic Compactifications},
arXiv:hep-th/0405014,JHEP0412:054,2004\\ 
Ron Donagi, Yang-Hui He, Burt Ovrut, Rene Reinbacher,
{\em Higgs Doublets, Split Multiplets and Heterotic $SU(3)_C x SU(2)_L x U(1)_Y$ Spectra},
arXiv:hep-th/0409291,Phys.Lett. B618 (2005) 259-264\\ 
Volker Braun, Burt A. Ovrut, Tony Pantev, Rene Reinbacher,
{\em Elliptic Calabi-Yau Threefolds with $Z_3 x Z_3$ Wilson Lines},
arXiv:hep-th/0410055, JHEP0412:062,2004\\ 
Evgeny I. Buchbinder, Burt A. Ovrut, Rene Reinbacher,
{\em Instanton Moduli in String Theory},
arXiv:hep-th/0410200, JHEP0504:008,2005\\ 
Ron Donagi, Yang-Hui He, Burt A. Ovrut, Rene Reinbacher,
{\em The Spectra of Heterotic Standard Model Vacua},
arXiv:hep-th/0411156, JHEP0506:070,2005\\
Volker Braun, Yang-Hui He, Burt A. Ovrut, Tony Pantev,
{\em A Heterotic Standard Model},
arXiv:hep-th/0501070, Phys.Lett.B618:252-258,2005\\ 
Volker Braun, Yang-Hui He, Burt A. Ovrut, Tony Pantev,
{\em A Standard Model from the E8 x E8 Heterotic Superstring},
arXiv:hep-th/0502155, JHEP 0506:039,2005\\
Volker Braun, Yang-Hui He, Burt A. Ovrut, Tony Pantev,
{\em Vector Bundle Extensions, Sheaf Cohomology, and the Heterotic Standard Model},
arXiv:hep-th/0505041, Adv.Theor.Math.Phys.10:4,2006\\ 
Volker Braun, Yang-Hui He, Burt A. Ovrut, Tony Pantev,
{\em Heterotic Standard Model Moduli},
arXiv:hep-th/0509051, JHEP 0601:025,2006\\ 
Volker Braun, Yang-Hui He, Burt A. Ovrut, Tony Pantev,
{\em Moduli Dependent mu-Terms in a Heterotic Standard Model},
arXiv:hep-th/0510142, JHEP 0603:006,2006\\ 
Volker Braun, Yang-Hui He, Burt A. Ovrut, Tony Pantev,
{\em The Exact MSSM Spectrum from String Theory},
arXiv:hep-th/0512177, JHEP 0605:043,2006\\ 
Volker Braun, Yang-Hui He, Burt A. Ovrut,
{\em Yukawa Couplings in Heterotic Standard Models},
arXiv:hep-th/0601204, JHEP 0604:019,2006\\ 
Volker Braun, Yang-Hui He, Burt A. Ovrut,
{\em Stability of the Minimal Heterotic Standard Model Bundle},
arXiv:hep-th/0602073, JHEP0606:032,2006\\ 
Volker Braun, Burt A. Ovrut,
{\em Stabilizing Moduli with a Positive Cosmological Constant in Heterotic M-Theory},
arXiv:hep-th/0603088, JHEP0607:035,2006\\ 
Volker Braun, Evgeny I. Buchbinder, Burt A.Ovrut,
{\em Dynamical SUSY Breaking in Heterotic M-Theory},
arXiv:hep-th/0606166, Phys.Lett.B639:566-570,2006\\ 
Volker Braun, Evgeny I. Buchbinder, Burt A. Ovrut,
{\em Towards Realizing Dynamical SUSY Breaking in Heterotic Model Building},
arXiv:hep-th/0606241, JHEP 0610:041,2006\\ 
James Gray, Andre Lukas, Burt Ovrut,
{\em Perturbative Anti-Brane Potentials in Heterotic M-theory},
arXiv:hep-th/0701025, Phys.Rev.D76:066007,2007\\ 
Volker Braun, Maximilian Kreuzer, Burt A. Ovrut, Emanuel Scheidegger,
{\em Worldsheet Instantons, Torsion Curves, and Non-Perturbative Superpotentials},
arXiv:hep-th/0703134, Phys.Lett.B649:334-341,2007\\
Volker Braun, Maximilian Kreuzer, Burt A. Ovrut, Emanuel Scheidegger,
{\em Worldsheet Instantons and Torsion Curves, Part A: Direct Computation},
arXiv:hep-th/0703182, JHEP0710:022,2007\\
Volker Braun, Maximilian Kreuzer, Burt A. Ovrut, Emanuel Scheidegger,
{\em Worldsheet Instantons and Torsion Curves, Part B: Mirror Symmetry},
arXiv:0704.0449, JHEP0710:023,2007\\ 
James Gray, André Lukas, Burt Ovrut,
{\em Flux, Gaugino Condensation and Anti-Branes in Heterotic M-theory},
arXiv:0709.2914, Phys.Rev.D76:126012,2007\\
Michael Ambroso, Volker Braun, Burt A. Ovrut,
{\em Two Higgs Pair Heterotic Vacua and Flavor-Changing Neutral Currents},
arXiv:0807.3319, JHEP0810:046,2008\\
Lara B. Anderson, James Gray, Andre Lukas, Burt Ovrut,
{\em The Edge Of Supersymmetry: Stability Walls in Heterotic Theory},
arXiv:0903.5088, Phys.Lett.B677:190-194,2009\\ 
Michael Ambroso, Burt Ovrut,
{\em The B-L/Electroweak Hierarchy in Heterotic String and M-Theory},
arXiv:0904.4509, JHEP 0910:011, 2009\\ 
Lara B. Anderson, James Gray, Andre Lukas, Burt Ovrut,
{\em Stability Walls in Heterotic Theories},
arXiv:0905.1748, JHEP 0909:026,2009\\ 
Michael Ambroso, Burt Ovrut, 
{\em The B-L/Electroweak Hierarchy in Smooth Heterotic Compactifications},
arXiv:0910.1129\\
Lara B. Anderson, James Gray, Burt Ovrut, 
{\em Yukawa Textures From Heterotic Stability Walls},
arXiv:1001.2317\\
T. Brelidze, B. Ovrut, 
{\em B-L Cosmic Strings in Heterotic Standard Models}, 
arXiv:1003.0234\\
Lara B. Anderson, Volker Braun, Robert L. Karp, Burt A. Ovrut, 
{\em Numerical Hermitian Yang-Mills Connections and Vector Bundle Stability in Heterotic Theories}
arXiv:1004.4399\\
Michael Ambroso, Burt A. Ovrut, 
{\em The Mass Spectra, Hierarchy and Cosmology of B-L MSSM Heterotic Compactifications},
arXiv:1005.5392\\
Lara B. Anderson, James Gray, Andre Lukas, Burt Ovrut,
{\em Stabilizing the Complex Structure in Heterotic Calabi-Yau Vacua},
arXiv:1010.0255, JHEP 1102:088,2011\\ 
Lara B. Anderson, James Gray, Burt Ovrut, 
{\em Transitions in the Web of Heterotic Vacua},
arXiv:1012.3179\\
Lara B. Anderson, James Gray, Andre Lukas, Burt Ovrut,
{\em Stabilizing All Geometric Moduli in Heterotic Calabi-Yau Vacua},
arXiv:1102.0011, Phys.Rev.D83:106011,2011\\
Lara B. Anderson, James Gray, Andre Lukas, Burt Ovrut, 
{\em The Atiyah Class and Complex Structure Stabilization in Heterotic Calabi-Yau Compactifications},
arXiv:1107.5076\\
Gottfried Curio, Ron Y. Donagi,
{\em Moduli in N=1 heterotic/F-theory duality},
arXiv:hep-th/9801057, Nucl.Phys. B518 (1998) 603-631\\ 
Ron Y. Donagi, 
{\em Taniguchi Lecture on Principal Bundles on Elliptic Fibrations},
arXiv:hep-th/9802094\\
Vincent Bouchard, Ron Donagi,
{\em An SU(5) Heterotic Standard Model}
arXiv:hep-th/0512149, Phys.Lett.B633:783-791,2006\\ 
Vincent Bouchard, Mirjam Cvetic, Ron Donagi,
{\em Tri-linear Couplings in an Heterotic Minimal Supersymmetric Standard Model},
arXiv:hep-th/0602096, Nucl.Phys.B745:62-83,2006\\ 
Ron Donagi, Rene Reinbacher, Shing-Tung-Yau, 
{\em Yukawa Couplings on Quintic Threefolds},
arXiv:hep-th/0605203\\
Vincent Bouchard, Ron Donagi,
{\em On a class of non-simply connected Calabi-Yau threefolds},
arXiv:0704.3096, Comm. Numb. Theor. Phys. 2 (2008) 1-61\\
Vincent Bouchard, Ron Donagi,
{\em On heterotic model constraints},
arXiv:0804.2096, JHEP 0808:060,2008\\ 
Anthony Bak, Vincent Bouchard, Ron Donagi,
{\em Exploring a new peak in the heterotic landscape},
arXiv:0811.1242, JHEP 06 (2010) 108, pp.1-31\\
Yang-Hui He, Maximilian Kreuzer, Seung-Joo Lee, Andre Lukas,
{\em Heterotic Bundles on Calabi-Yau Manifolds with Small Picard Number}
arXiv:1108.1031\\
Yang-Hui He,
{\em An Algorithmic Approach to Heterotic String Phenomenology},
arXiv:1001.2419, Mod. Phys. Lett. A, Vol. 25, No. 2 (2010) pp. 79-90\\ 
Lara B. Anderson, James Gray, Yang-Hui He, Andre Lukas,
{\em Exploring Positive Monad Bundles And A New Heterotic Standard Model},
arXiv:0911.1569\\ 
Yang-Hui He, Seung-Joo Lee, Andre Lukas,
{\em Heterotic Models from Vector Bundles on Toric Calabi-Yau Manifolds},
arXiv:0911.0865, JHEP 1005:071,2010\\ 
Lara B. Anderson, James Gray, Dan Grayson, Yang-Hui He, Andre Lukas,
{\em Yukawa Couplings in Heterotic Compactification},
arXiv:0904.2186, Commun.Math.Phys.297:95-127,2010\\ 
Maxime Gabella, Yang-Hui He, Andre Lukas,
{\em An Abundance of Heterotic Vacua},
arXiv:0808.2142, JHEP0812:027,2008\\
Lara B. Anderson, Yang-Hui He, Andre Lukas,
{\em Monad Bundles in Heterotic String Compactifications},
arXiv:0805.2875, JHEP 0807:104,2008\\
Lara B. Anderson, Yang-Hui He, Andre Lukas,
{\em Heterotic Compactification, An Algorithmic Approach},
arXiv:hep-th/0702210, JHEP 0707:049,2007\\ 
Bjorn Andreas, Gottfried Curio, 
{\em On the Existence of Stable bundles with prescribed Chern classes on Calabi-Yau threefolds},
arXiv:1104.3435\\
Bjorn Andreas, Gottfried Curio, 
{\em Spectral Bundles and the DRY-Conjecture},
arXiv:1012.3858\\
Bjorn Andreas, Gottfried Curio,
{\em On possible Chern Classes of stable Bundles on Calabi-Yau threefolds},
arXiv:1010.1644, J.Geom.Phys.61:1378-1384,2011\\ 
Gottfried Curio, 
{\em On the Heterotic World-sheet Instanton Superpotential and its individual Contributions},
arXiv:1006.5568\\
Gottfried Curio,
{\em Perspectives on Pfaffians of Heterotic World-sheet Instantons},
arXiv:0904.2738, JHEP 0909:131,2009\\ 
Gottfried Curio,
{\em World-sheet Instanton Superpotentials in Heterotic String theory and their Moduli Dependence},
arXiv:0810.3087, JHEP 0909:125,2009\\ 
Bjorn Andreas, Gottfried Curio,
{\em Deformations of Bundles and the Standard Model},
arXiv:0706.1158, Phys.Lett.B655:290-293,2007\\ 
Bjorn Andreas, Gottfried Curio,
{\em Extension Bundles and the Standard Model},
arXiv:hep-th/0703210, JHEP0707:053,2007\\ 
Bjorn Andreas, Gottfried Curio,
{\em Heterotic Models without Fivebranes}
arXiv:hep-th/0611309, J.Geom.Phys.57:2136-2145,2007\\
Bjorn Andreas, Gottfried Curio, 
{\em Invariant Bundles on $B$-fibered Calabi-Yau Spaces and the Standard Model},
arXiv:hep-th/0602247\\
Gottfried Curio,
{\em Standard Model bundles of the heterotic string},
arXiv:hep-th/0412182 Int.J.Mod.Phys. A21 (2006) 1261-1282\\
Bjorn Andreas, Gottfried Curio, Albrecht Klemm,
{\em Towards the Standard Model spectrum from elliptic Calabi-Yau},
arXiv:hep-th/9903052, Int.J.Mod.Phys. A19 (2004) 1987\\
G. Curio, 
{\em Chiral matter and transitions in heterotic string models},
arXiv:hep-th/9803224, Phys.Lett. B435 (1998) 39-48.

\end{enumerate}
\end{document}